\documentclass[acmsmall, screen, authorversion, nonacm]{acmart}

\usepackage[utf8]{inputenc}

\usepackage[references]{agda}
\usepackage{xcolor}
\usepackage{cleveref}
\usepackage{stmaryrd}

\setcitestyle{nosort}
\usepackage{newunicodechar}
\newunicodechar{𝔽}{\ensuremath{\mathbb{F}}}
\newunicodechar{𝕃}{\ensuremath{\mathbb{L}}}
\newunicodechar{ℙ}{\ensuremath{\mathbb{P}}}
\newunicodechar{ℂ}{\ensuremath{\mathbb{C}}}
\newunicodechar{＠}{\ensuremath{@}}
\newunicodechar{≟}{\ensuremath{=^?}}
\newunicodechar{₁}{\ensuremath{_1}}
\newunicodechar{₂}{\ensuremath{_2}}
\newunicodechar{◁}{\ensuremath{\lhd}}
\newunicodechar{⟦}{\ensuremath{\llbracket}}
\newunicodechar{⟧}{\ensuremath{\rrbracket}}
\newunicodechar{▷}{\ensuremath{\rhd}}
\newunicodechar{⇒}{\ensuremath{\Rightarrow}}
\newunicodechar{◇}{\ensuremath{\diamond}}
\newunicodechar{ⁿ}{\ensuremath{^n}}
\newunicodechar{ᶜ}{\ensuremath{^c}}
\newunicodechar{∀}{\ensuremath{\forall}}
\newunicodechar{Σ}{\ensuremath{\Sigma}}
\newunicodechar{ω}{\ensuremath{\omega}}
\newunicodechar{∈}{\ensuremath{\in}}
\newunicodechar{∘}{\ensuremath{\circ}}
\newunicodechar{′}{{'}}
\newunicodechar{ℓ}{\ensuremath{\ell}}
\newunicodechar{⊤}{\ensuremath{\top}}
\newunicodechar{λ}{\ensuremath{\lambda}}

\newcommand{\locally}[2]{\ensuremath{#1 \mathbin{\color{orange}\rhd} #2}}
\newcommand{\comm}[3]{\ensuremath{#1 \mathbin{\color{orange}\Rightarrow} #2 \mathbin{\color{orange}\square} #3}}

\iftrue % use \iftrue / \iffalse to turn on/off comments
  \newcommand{\newcommenter}[3]{%
    \newcommand{#1}[1]{%
      \textcolor{#2}{\small\textsf{[{#3}: {##1}]}}%
    }%
  }
\else
  \newcommand{\newcommenter}[3]{\newcommand{#1}[1]{}}
\fi

\newcommenter{\gs}{blue}{GS}
\newcommenter{\lk}{red}{LK}

\title{Toward Verified Library-Level Choreographic Programming with Algebraic Effects}

\author{Gan Shen}
\orcid{0009-0006-0947-9531}
\affiliation{%
  \institution{University of California, Santa Cruz}
  \country{USA}}

\author{Lindsey Kuper}
\orcid{0000-0002-1374-7715}
\affiliation{%
  \institution{University of California, Santa Cruz}
  \country{USA}}

\begin{abstract}
  Choreographic programming (CP) is a paradigm for programming distributed applications
  as single, unified programs, called \emph{choreographies}, that are then compiled to node-local programs via \emph{endpoint projection} (EPP).
  Recently, \emph{library-level} CP frameworks have emerged, in which choreographies and EPP are expressed as constructs in an existing host language.
  So far, however, library-level CP lacks a solid theoretical foundation.

  In this paper, we propose modeling library-level CP using \emph{algebraic effects}, an abstraction that generalizes the approach taken by existing CP libraries.
  Algebraic effects let us define choreographies as computations with user-defined effects and EPP as location-specific effect handlers.
  Algebraic effects also lend themselves to reasoning about correctness properties, such as soundness and completeness of EPP.
  We present a prototype of a library-level CP framework based on algebraic effects, implemented in the Agda proof assistant, and discuss our ongoing work on leveraging the algebraic-effects-based approach to prove the correctness of our library-level CP implementation.
\end{abstract}

\begin{document}

\maketitle

\section{Introduction}

Choreographic programming~(CP)~\citep{montesi-2013, montesi-2023} is a paradigm for programming distributed applications that run on multiple nodes.
In CP the programmer writes one, unified program, called a \emph{choreography}, that is then compiled to individual programs for each node via a compilation step called \emph{endpoint projection} (EPP).
For example, the following choreography describes a distributed data processing pipeline, involving nodes Alice, Bob, and Carol, which respectively run functions $f$, $g$, and $h$ on their input:
\begin{equation*}
  \begin{split}
    x \leftarrow \; & \locally{\mathrm{Alice}}{\mathsf{getInput}} \\
    y \leftarrow \; & \comm{\mathrm{Alice}}{\mathrm{Bob}}{f(x)} \\
    z \leftarrow \; & \comm{\mathrm{Bob}}{\mathrm{Carol}}{g(y)} \\
    w \leftarrow \; & \comm{\mathrm{Carol}}{\mathrm{Alice}}{h(z)} \\
                    & \locally{\mathrm{Alice}}{\mathsf{showResults}(w)} \\
  \end{split}
\end{equation*}
Here, we use $\leftarrow$ for variable bindings;
$\locally{\mathrm{Alice}}{t}$ denotes a local computation $t$ at Alice;
and $\comm{\mathrm{Alice}}{\mathrm{Bob}}{t}$ denotes communication from Alice to Bob with message $t$.
In this choreography, Alice first gets some input locally, processes it with $f$, and passes the result to Bob, who processes it with $g$ and passes the result to Carol, who processes it with $h$ and passes it back to Alice to be displayed to the user.
To get an executable program for each node, we can apply endpoint projection to the choreography, resulting in individual programs for Alice, Bob, and Carol:
\hspace{-1cm}
\[
\begin{minipage}{.35\textwidth}
  \begin{equation*}
    \begin{split}
      x \leftarrow \; & \mathsf{getInput} \\
                      & \mathsf{send}(\mathrm{Bob}, f(x)) \\
      w \leftarrow \; & \mathsf{recv}(\mathrm{Carol}) \\
                      & \mathsf{showResults}(w)
    \end{split}
  \end{equation*}
\end{minipage}
\hfill\vline\hfill
\begin{minipage}{.3\textwidth}
  \begin{equation*}
    \begin{split}
      y \leftarrow \; & \mathsf{recv}(\mathrm{Alice}) \\
                      & \mathsf{send}(\mathrm{Carol}, g(y)) \\
    \end{split}
  \end{equation*}
\end{minipage}
\hfill\vline\hfill
\begin{minipage}{.3\textwidth}
  \begin{equation*}
    \begin{split}
      z \leftarrow \; & \mathsf{recv}(\mathrm{Bob}) \\
                      & \mathsf{send}(\mathrm{Alice}, h(z)) \\
    \end{split}
  \end{equation*}
\end{minipage}
\]
A correct CP language guarantees soundness and completeness of EPP, which further implies that the collection of projected programs is deadlock-free when running together.
Existing research places CP on a solid theoretical foundation~\citep{montesi-2013, cruzfilipe-2020, cruzfilipe-2022, hirsch-2022},
which has informed the design of practical, full-featured standalone CP languages such as Choral~\citep{giallorenzo-2024}.

Recent work has introduced \emph{library-level} CP languages~\citep{shen-2023, kashiwa-2023}, in which choreographies and EPP are completely expressed as constructs in an existing host language.  For example, HasChor~\citep{shen-2023}, implements support for CP by means of a domain-specific language embedded in Haskell.
In HasChor, choreographies are monadic computations in which choreographic operators such as $\locally{\_}{\_}$ and $\comm{\_}{\_}{\_}$ may be used, and EPP is carried out by means of \emph{dynamic interpretation} of choreographies at run time.  The recently proposed ChoRus library for choreographic programming in Rust~\citep{kashiwa-2023} takes a similarly dynamic approach.
Library-level CP frameworks have the potential to improve the accessibility and practicality of CP by integrating it into general-purpose programming languages.
However, there are no proofs of correctness of EPP for library-level CP frameworks. Indeed, it is unclear to what extent the established theory of CP is applicable in the setting of library-level CP.

To close this gap, in this paper we propose \emph{algebraic effects}~\citep{plotkin-2003, plotkin-2013} as a foundational approach for implementing and verifying library-level CP.
Algebraic effects provide an abstraction that generalizes existing approaches to library-level CP.
In particular, they allow us to define choreographies as computations with user-defined effects and EPP as location-specific effect handlers.
Algebraic effects also lend themselves to proofs of correctness.
They provide abstract syntax trees for choreographies in which CP-specific effects and control flows are manifest, enabling reasoning about them.
Furthermore, given that algebraic effects are ``going mainstream''~\citep{dagstuhl-alg-effects-report}, with efficient implementations now available in languages such as OCaml~\citep{sivaramakrishnan-ocaml-effect-handlers} and Koka~\citep{leijen-2017}, we believe our proposed approach would make library-level CP less ad-hoc and bring it to a broader audience.

In the rest of the paper, we set up a framework for programming with algebraic effects in Agda (\Cref{sec:alg-eff-agda}), which we then use to implement a prototype library-level CP framework (\Cref{sec:cp-alg-eff}).  Finally, we discuss our ongoing work on leveraging our approach to prove the correctness of our library-level CP implementation (\Cref{sec:next}).  This paper is a literate Agda program.

\section{A Minimal Algebraic Effects Framework in Agda}
\label{sec:alg-eff-agda}

In this section, we define a minimal algebraic effects framework in Agda, which we will use to implement CP in the next section.
No prior knowledge of algebraic effects is assumed.
We introduce each concept first from a mathematical perspective and then relate it to programming.
Due to lack of space, we do not include any examples in this section, but the next section can be seen as a demonstration of the framework.
Our presentation is influenced by \citet{bauer-2019} and \citet{kidney-2023}, and we refer the reader to them for a comprehensive introduction to algebraic effects.

\begin{code}[hide]%
\>[0]\AgdaKeyword{open}\AgdaSpace{}%
\AgdaKeyword{import}\AgdaSpace{}%
\AgdaModule{Data.Product}\AgdaSpace{}%
\AgdaKeyword{using}\AgdaSpace{}%
\AgdaSymbol{(}\AgdaRecord{Σ}\AgdaSymbol{;}\AgdaSpace{}%
\AgdaFunction{Σ-syntax}\AgdaSymbol{;}\AgdaSpace{}%
\AgdaOperator{\AgdaInductiveConstructor{\AgdaUnderscore{},\AgdaUnderscore{}}}\AgdaSymbol{)}\<%
\\
\>[0]\AgdaKeyword{open}\AgdaSpace{}%
\AgdaKeyword{import}\AgdaSpace{}%
\AgdaModule{Effect.Applicative}\AgdaSpace{}%
\AgdaKeyword{using}\AgdaSpace{}%
\AgdaSymbol{(}\AgdaRecord{RawApplicative}\AgdaSymbol{)}\<%
\\
\>[0]\AgdaKeyword{open}\AgdaSpace{}%
\AgdaKeyword{import}\AgdaSpace{}%
\AgdaModule{Effect.Functor}\AgdaSpace{}%
\AgdaKeyword{using}\AgdaSpace{}%
\AgdaSymbol{(}\AgdaRecord{RawFunctor}\AgdaSymbol{)}\<%
\\
\>[0]\AgdaKeyword{open}\AgdaSpace{}%
\AgdaKeyword{import}\AgdaSpace{}%
\AgdaModule{Effect.Monad}\AgdaSpace{}%
\AgdaKeyword{using}\AgdaSpace{}%
\AgdaSymbol{(}\AgdaRecord{RawMonad}\AgdaSymbol{)}\<%
\\
\>[0]\AgdaKeyword{open}\AgdaSpace{}%
\AgdaKeyword{import}\AgdaSpace{}%
\AgdaModule{Function}\AgdaSpace{}%
\AgdaKeyword{using}\AgdaSpace{}%
\AgdaSymbol{(}\AgdaOperator{\AgdaFunction{\AgdaUnderscore{}∘\AgdaUnderscore{}}}\AgdaSymbol{;}\AgdaSpace{}%
\AgdaOperator{\AgdaFunction{\AgdaUnderscore{}∘′\AgdaUnderscore{}}}\AgdaSymbol{;}\AgdaSpace{}%
\AgdaOperator{\AgdaFunction{\AgdaUnderscore{}\$′\AgdaUnderscore{}}}\AgdaSymbol{)}\<%
\\
\>[0]\AgdaKeyword{open}\AgdaSpace{}%
\AgdaKeyword{import}\AgdaSpace{}%
\AgdaModule{Level}\AgdaSpace{}%
\AgdaKeyword{using}\AgdaSpace{}%
\AgdaSymbol{(}\AgdaPostulate{Level}\AgdaSymbol{)}\<%
\\
\\[\AgdaEmptyExtraSkip]%
\>[0]\AgdaKeyword{infix}\AgdaSpace{}%
\AgdaNumber{21}\AgdaSpace{}%
\AgdaOperator{\AgdaFunction{⟦\AgdaUnderscore{}⟧}}\<%
\end{code}

\begin{code}[hide]%
\>[0]\AgdaComment{--\ the\ standard\ library\ `mkRawMonad`\ is\ not\ general\ enough\ for\ our\ use}\<%
\\
\>[0]\AgdaComment{--\ so\ we\ define\ our\ own\ version}\<%
\\
\>[0]\AgdaKeyword{module}\AgdaSpace{}%
\AgdaModule{\AgdaUnderscore{}}\AgdaSpace{}%
\AgdaKeyword{where}\<%
\\
\\[\AgdaEmptyExtraSkip]%
\>[0][@{}l@{\AgdaIndent{0}}]%
\>[2]\AgdaKeyword{private}\<%
\\
\>[2][@{}l@{\AgdaIndent{0}}]%
\>[4]\AgdaKeyword{variable}\<%
\\
\>[4][@{}l@{\AgdaIndent{0}}]%
\>[6]\AgdaGeneralizable{f}\AgdaSpace{}%
\AgdaGeneralizable{g}\AgdaSpace{}%
\AgdaSymbol{:}\AgdaSpace{}%
\AgdaPostulate{Level}\<%
\\
\\[\AgdaEmptyExtraSkip]%
\>[2]\AgdaKeyword{open}\AgdaSpace{}%
\AgdaModule{RawFunctor}\<%
\\
\>[2]\AgdaKeyword{open}\AgdaSpace{}%
\AgdaModule{RawApplicative}\<%
\\
\>[2]\AgdaKeyword{open}\AgdaSpace{}%
\AgdaModule{RawMonad}\<%
\\
\\[\AgdaEmptyExtraSkip]%
\>[2]\AgdaFunction{mkRawApplicative}\AgdaSpace{}%
\AgdaSymbol{:}\<%
\\
\>[2][@{}l@{\AgdaIndent{0}}]%
\>[4]\AgdaSymbol{(}\AgdaBound{F}\AgdaSpace{}%
\AgdaSymbol{:}\AgdaSpace{}%
\AgdaPrimitive{Set}\AgdaSpace{}%
\AgdaGeneralizable{f}\AgdaSpace{}%
\AgdaSymbol{→}\AgdaSpace{}%
\AgdaPrimitive{Set}\AgdaSpace{}%
\AgdaGeneralizable{g}\AgdaSymbol{)}\AgdaSpace{}%
\AgdaSymbol{→}\<%
\\
\>[4]\AgdaSymbol{(}\AgdaBound{pure}\AgdaSpace{}%
\AgdaSymbol{:}\AgdaSpace{}%
\AgdaSymbol{∀}\AgdaSpace{}%
\AgdaSymbol{\{}\AgdaBound{A}\AgdaSymbol{\}}\AgdaSpace{}%
\AgdaSymbol{→}\AgdaSpace{}%
\AgdaBound{A}\AgdaSpace{}%
\AgdaSymbol{→}\AgdaSpace{}%
\AgdaBound{F}\AgdaSpace{}%
\AgdaBound{A}\AgdaSymbol{)}\AgdaSpace{}%
\AgdaSymbol{→}\<%
\\
\>[4]\AgdaSymbol{(}\AgdaBound{app}\AgdaSpace{}%
\AgdaSymbol{:}\AgdaSpace{}%
\AgdaSymbol{∀}\AgdaSpace{}%
\AgdaSymbol{\{}\AgdaBound{A}\AgdaSpace{}%
\AgdaBound{B}\AgdaSymbol{\}}\AgdaSpace{}%
\AgdaSymbol{→}\AgdaSpace{}%
\AgdaBound{F}\AgdaSpace{}%
\AgdaSymbol{(}\AgdaBound{A}\AgdaSpace{}%
\AgdaSymbol{→}\AgdaSpace{}%
\AgdaBound{B}\AgdaSymbol{)}\AgdaSpace{}%
\AgdaSymbol{→}\AgdaSpace{}%
\AgdaBound{F}\AgdaSpace{}%
\AgdaBound{A}\AgdaSpace{}%
\AgdaSymbol{→}\AgdaSpace{}%
\AgdaBound{F}\AgdaSpace{}%
\AgdaBound{B}\AgdaSymbol{)}\AgdaSpace{}%
\AgdaSymbol{→}\<%
\\
\>[4]\AgdaRecord{RawApplicative}\AgdaSpace{}%
\AgdaBound{F}\<%
\\
\>[2]\AgdaFunction{mkRawApplicative}\AgdaSpace{}%
\AgdaBound{F}\AgdaSpace{}%
\AgdaBound{pure}\AgdaSpace{}%
\AgdaBound{app}\AgdaSpace{}%
\AgdaSymbol{.}\AgdaField{rawFunctor}\AgdaSpace{}%
\AgdaSymbol{.}\AgdaOperator{\AgdaField{\AgdaUnderscore{}<\$>\AgdaUnderscore{}}}\AgdaSpace{}%
\AgdaSymbol{=}\AgdaSpace{}%
\AgdaBound{app}\AgdaSpace{}%
\AgdaOperator{\AgdaFunction{∘′}}\AgdaSpace{}%
\AgdaBound{pure}\<%
\\
\>[2]\AgdaFunction{mkRawApplicative}\AgdaSpace{}%
\AgdaBound{F}\AgdaSpace{}%
\AgdaBound{pure}\AgdaSpace{}%
\AgdaBound{app}\AgdaSpace{}%
\AgdaSymbol{.}\AgdaField{pure}\AgdaSpace{}%
\AgdaSymbol{=}\AgdaSpace{}%
\AgdaBound{pure}\<%
\\
\>[2]\AgdaFunction{mkRawApplicative}\AgdaSpace{}%
\AgdaBound{F}\AgdaSpace{}%
\AgdaBound{pure}\AgdaSpace{}%
\AgdaBound{app}\AgdaSpace{}%
\AgdaSymbol{.}\AgdaOperator{\AgdaField{\AgdaUnderscore{}<*>\AgdaUnderscore{}}}\AgdaSpace{}%
\AgdaSymbol{=}\AgdaSpace{}%
\AgdaBound{app}\<%
\\
\\[\AgdaEmptyExtraSkip]%
\>[2]\AgdaFunction{mkRawMonad}\AgdaSpace{}%
\AgdaSymbol{:}\<%
\\
\>[2][@{}l@{\AgdaIndent{0}}]%
\>[4]\AgdaSymbol{(}\AgdaBound{F}\AgdaSpace{}%
\AgdaSymbol{:}\AgdaSpace{}%
\AgdaPrimitive{Set}\AgdaSpace{}%
\AgdaGeneralizable{f}\AgdaSpace{}%
\AgdaSymbol{→}\AgdaSpace{}%
\AgdaPrimitive{Set}\AgdaSpace{}%
\AgdaGeneralizable{g}\AgdaSymbol{)}\AgdaSpace{}%
\AgdaSymbol{→}\<%
\\
\>[4]\AgdaSymbol{(}\AgdaBound{pure}\AgdaSpace{}%
\AgdaSymbol{:}\AgdaSpace{}%
\AgdaSymbol{∀}\AgdaSpace{}%
\AgdaSymbol{\{}\AgdaBound{A}\AgdaSymbol{\}}\AgdaSpace{}%
\AgdaSymbol{→}\AgdaSpace{}%
\AgdaBound{A}\AgdaSpace{}%
\AgdaSymbol{→}\AgdaSpace{}%
\AgdaBound{F}\AgdaSpace{}%
\AgdaBound{A}\AgdaSymbol{)}\AgdaSpace{}%
\AgdaSymbol{→}\<%
\\
\>[4]\AgdaSymbol{(}\AgdaBound{bind}\AgdaSpace{}%
\AgdaSymbol{:}\AgdaSpace{}%
\AgdaSymbol{∀}\AgdaSpace{}%
\AgdaSymbol{\{}\AgdaBound{A}\AgdaSpace{}%
\AgdaBound{B}\AgdaSymbol{\}}\AgdaSpace{}%
\AgdaSymbol{→}\AgdaSpace{}%
\AgdaBound{F}\AgdaSpace{}%
\AgdaBound{A}\AgdaSpace{}%
\AgdaSymbol{→}\AgdaSpace{}%
\AgdaSymbol{(}\AgdaBound{A}\AgdaSpace{}%
\AgdaSymbol{→}\AgdaSpace{}%
\AgdaBound{F}\AgdaSpace{}%
\AgdaBound{B}\AgdaSymbol{)}\AgdaSpace{}%
\AgdaSymbol{→}\AgdaSpace{}%
\AgdaBound{F}\AgdaSpace{}%
\AgdaBound{B}\AgdaSymbol{)}\AgdaSpace{}%
\AgdaSymbol{→}\<%
\\
\>[4]\AgdaRecord{RawMonad}\AgdaSpace{}%
\AgdaBound{F}\<%
\\
\>[2]\AgdaFunction{mkRawMonad}\AgdaSpace{}%
\AgdaBound{F}\AgdaSpace{}%
\AgdaBound{pure}\AgdaSpace{}%
\AgdaOperator{\AgdaBound{\AgdaUnderscore{}>>=\AgdaUnderscore{}}}\AgdaSpace{}%
\AgdaSymbol{.}\AgdaField{rawApplicative}\AgdaSpace{}%
\AgdaSymbol{=}\<%
\\
\>[2][@{}l@{\AgdaIndent{0}}]%
\>[4]\AgdaFunction{mkRawApplicative}\AgdaSpace{}%
\AgdaSymbol{\AgdaUnderscore{}}\AgdaSpace{}%
\AgdaBound{pure}\AgdaSpace{}%
\AgdaOperator{\AgdaFunction{\$′}}\AgdaSpace{}%
\AgdaSymbol{λ}\AgdaSpace{}%
\AgdaBound{mf}\AgdaSpace{}%
\AgdaBound{mx}\AgdaSpace{}%
\AgdaSymbol{→}\AgdaSpace{}%
\AgdaKeyword{do}\<%
\\
\>[4][@{}l@{\AgdaIndent{0}}]%
\>[6]\AgdaBound{f}\AgdaSpace{}%
\AgdaOperator{\AgdaBound{←}}\AgdaSpace{}%
\AgdaBound{mf}\<%
\\
\>[6]\AgdaBound{x}\AgdaSpace{}%
\AgdaOperator{\AgdaBound{←}}\AgdaSpace{}%
\AgdaBound{mx}\<%
\\
\>[6]\AgdaBound{pure}\AgdaSpace{}%
\AgdaSymbol{(}\AgdaBound{f}\AgdaSpace{}%
\AgdaBound{x}\AgdaSymbol{)}\<%
\\
\>[2]\AgdaFunction{mkRawMonad}\AgdaSpace{}%
\AgdaBound{F}\AgdaSpace{}%
\AgdaBound{pure}\AgdaSpace{}%
\AgdaOperator{\AgdaBound{\AgdaUnderscore{}>>=\AgdaUnderscore{}}}\AgdaSpace{}%
\AgdaSymbol{.}\AgdaOperator{\AgdaField{\AgdaUnderscore{}>>=\AgdaUnderscore{}}}\AgdaSpace{}%
\AgdaSymbol{=}\AgdaSpace{}%
\AgdaOperator{\AgdaBound{\AgdaUnderscore{}>>=\AgdaUnderscore{}}}\<%
\end{code}

\subsection{Signatures and Algebras}

A signature \AgdaRecord{Sig} specifies the equipped operations of an algebra, which includes a type \AgdaField{Op} of operations and a function \AgdaField{Arity} giving the number of arguments (represented as the cardinality of a type) of each operation:\footnote{Agda uses an infinite hierarchy of universes where $\AgdaPrimitive{Set} : \AgdaPrimitive{Set}_1 : ... : \AgdaPrimitive{Set}_n : \AgdaPrimitive{Set}_{n+1}$ to avoid paradoxes.
For ease of presentation in this paper, we overconstrain the universe of \AgdaField{Op} to be \AgdaPrimitive{Set₁} (similarily for \AgdaField{Arity}).
The actual implementation is universe-polymorphic,
but readers can safely ignore the universe hierarchy without missing the key point of the paper.}
\begin{center}\begin{code}%
\>[0]\AgdaKeyword{record}\AgdaSpace{}%
\AgdaRecord{Sig}\AgdaSpace{}%
\AgdaSymbol{:}\AgdaSpace{}%
\AgdaPrimitive{Set₂}\AgdaSpace{}%
\AgdaKeyword{where}\<%
\\
\>[0][@{}l@{\AgdaIndent{0}}]%
\>[2]\AgdaKeyword{constructor}\AgdaSpace{}%
\AgdaOperator{\AgdaInductiveConstructor{\AgdaUnderscore{}◁\AgdaUnderscore{}}}\<%
\\
\>[2]\AgdaKeyword{field}\<%
\\
\>[2][@{}l@{\AgdaIndent{0}}]%
\>[4]\AgdaField{Op}\AgdaSpace{}%
\AgdaSymbol{:}\AgdaSpace{}%
\AgdaPrimitive{Set₁}\<%
\\
\>[4]\AgdaField{Arity}\AgdaSpace{}%
\AgdaSymbol{:}\AgdaSpace{}%
\AgdaField{Op}\AgdaSpace{}%
\AgdaSymbol{→}\AgdaSpace{}%
\AgdaPrimitive{Set}\<%
\end{code}\end{center}
\begin{code}[hide]%
\>[0]\AgdaKeyword{open}\AgdaSpace{}%
\AgdaModule{Sig}\<%
\end{code}
A set $X$ that implements the operations of a signature $\mathbb{F}$ is called an $\mathbb{F}$-algebra.
An $\mathbb{F}$-algebra on the carrier set $X$, written as $\mathbb{F}$ \AgdaFunction{-Alg[} $X$ \AgdaFunction{]} in Agda, is a function of type \AgdaFunction{⟦} 𝔽 \AgdaFunction{⟧} $X \rightarrow X$:
\[
\begin{minipage}{.5\textwidth}
\begin{center}\begin{code}%
\>[0]\AgdaOperator{\AgdaFunction{⟦\AgdaUnderscore{}⟧}}\AgdaSpace{}%
\AgdaSymbol{:}\AgdaSpace{}%
\AgdaRecord{Sig}\AgdaSpace{}%
\AgdaSymbol{→}\AgdaSpace{}%
\AgdaPrimitive{Set₁}\AgdaSpace{}%
\AgdaSymbol{→}\AgdaSpace{}%
\AgdaPrimitive{Set₁}\<%
\\
\>[0]\AgdaOperator{\AgdaFunction{⟦}}\AgdaSpace{}%
\AgdaBound{Op}\AgdaSpace{}%
\AgdaOperator{\AgdaInductiveConstructor{◁}}\AgdaSpace{}%
\AgdaBound{Ar}\AgdaSpace{}%
\AgdaOperator{\AgdaFunction{⟧}}\AgdaSpace{}%
\AgdaBound{X}\AgdaSpace{}%
\AgdaSymbol{=}\AgdaSpace{}%
\AgdaFunction{Σ[}\AgdaSpace{}%
\AgdaBound{o}\AgdaSpace{}%
\AgdaFunction{∈}\AgdaSpace{}%
\AgdaBound{Op}\AgdaSpace{}%
\AgdaFunction{]}\AgdaSpace{}%
\AgdaSymbol{(}\AgdaBound{Ar}\AgdaSpace{}%
\AgdaBound{o}\AgdaSpace{}%
\AgdaSymbol{→}\AgdaSpace{}%
\AgdaBound{X}\AgdaSymbol{)}\<%
\end{code}\end{center}
\end{minipage}
\begin{minipage}{.5\textwidth}
\begin{center}\begin{code}%
\>[0]\AgdaOperator{\AgdaFunction{\AgdaUnderscore{}-Alg[\AgdaUnderscore{}]}}\AgdaSpace{}%
\AgdaSymbol{:}\AgdaSpace{}%
\AgdaRecord{Sig}\AgdaSpace{}%
\AgdaSymbol{→}\AgdaSpace{}%
\AgdaPrimitive{Set₁}\AgdaSpace{}%
\AgdaSymbol{→}\AgdaSpace{}%
\AgdaPrimitive{Set₁}\<%
\\
\>[0]\AgdaBound{𝔽}\AgdaSpace{}%
\AgdaOperator{\AgdaFunction{-Alg[}}\AgdaSpace{}%
\AgdaBound{X}\AgdaSpace{}%
\AgdaOperator{\AgdaFunction{]}}\AgdaSpace{}%
\AgdaSymbol{=}\AgdaSpace{}%
\AgdaOperator{\AgdaFunction{⟦}}\AgdaSpace{}%
\AgdaBound{𝔽}\AgdaSpace{}%
\AgdaOperator{\AgdaFunction{⟧}}\AgdaSpace{}%
\AgdaBound{X}\AgdaSpace{}%
\AgdaSymbol{→}\AgdaSpace{}%
\AgdaBound{X}\<%
\end{code}\end{center}
\end{minipage}
\]
\AgdaFunction{⟦} 𝔽 \AgdaFunction{⟧} $X$ denotes an operation paired with its arity number of elements from the carrier set --- a fully applied operation.
\AgdaFunction{⟦} 𝔽 \AgdaFunction{⟧} $X \rightarrow X$ allows us to make a new element out of a fully applied operation, the very nature of an algebra.

In programming, we can view an effectful operation as giving rise to an algebra, with the allowed effects being the signature, the result of an effect being the arity of the operation, and a carrier set of such an algebra being a functional model of the effectful computation.

\subsection{The Free Algebra}

Among all the algebras of a signature $\mathbb{F}$, we are particularly interested in one called the \emph{free algebra}.
Rather than performing operations in the carrier set, the free algebra merely records them as a data type, which we call \AgdaDatatype{Term}:
\begin{center}\begin{code}%
\>[0]\AgdaKeyword{data}\AgdaSpace{}%
\AgdaDatatype{Term}\AgdaSpace{}%
\AgdaSymbol{(}\AgdaBound{𝔽}\AgdaSpace{}%
\AgdaSymbol{:}\AgdaSpace{}%
\AgdaRecord{Sig}\AgdaSymbol{)}\AgdaSpace{}%
\AgdaSymbol{(}\AgdaBound{A}\AgdaSpace{}%
\AgdaSymbol{:}\AgdaSpace{}%
\AgdaPrimitive{Set}\AgdaSymbol{)}\AgdaSpace{}%
\AgdaSymbol{:}\AgdaSpace{}%
\AgdaPrimitive{Set₁}\AgdaSpace{}%
\AgdaKeyword{where}\<%
\\
\>[0][@{}l@{\AgdaIndent{0}}]%
\>[2]\AgdaInductiveConstructor{var}\AgdaSpace{}%
\AgdaSymbol{:}\AgdaSpace{}%
\AgdaBound{A}\AgdaSpace{}%
\AgdaSymbol{→}\AgdaSpace{}%
\AgdaDatatype{Term}\AgdaSpace{}%
\AgdaBound{𝔽}\AgdaSpace{}%
\AgdaBound{A}\<%
\\
\>[2]\AgdaInductiveConstructor{op}\AgdaSpace{}%
\AgdaSymbol{:}\AgdaSpace{}%
\AgdaOperator{\AgdaFunction{⟦}}\AgdaSpace{}%
\AgdaBound{𝔽}\AgdaSpace{}%
\AgdaOperator{\AgdaFunction{⟧}}\AgdaSpace{}%
\AgdaSymbol{(}\AgdaDatatype{Term}\AgdaSpace{}%
\AgdaBound{𝔽}\AgdaSpace{}%
\AgdaBound{A}\AgdaSymbol{)}\AgdaSpace{}%
\AgdaSymbol{→}\AgdaSpace{}%
\AgdaDatatype{Term}\AgdaSpace{}%
\AgdaBound{𝔽}\AgdaSpace{}%
\AgdaBound{A}\<%
\end{code}\end{center}
The \AgdaInductiveConstructor{var} constructor denotes a variable drawn from some set $A$.
The \AgdaInductiveConstructor{op} constructor denotes a fully applied operation to some other terms.

In programming, \AgdaDatatype{Term}s correspond to programs where effects are left uninterpreted, with variables being pure computations and operations being effectful computations.
\AgdaDatatype{Term}s also form a monad (it is actually the free monad), which allows us to chain them together:
\begin{center}\begin{code}%
\>[0]\AgdaFunction{return}\AgdaSpace{}%
\AgdaSymbol{:}\AgdaSpace{}%
\AgdaSymbol{∀}\AgdaSpace{}%
\AgdaSymbol{\{}\AgdaBound{𝔽}\AgdaSymbol{\}}\AgdaSpace{}%
\AgdaSymbol{\{}\AgdaBound{A}\AgdaSymbol{\}}\AgdaSpace{}%
\AgdaSymbol{→}\AgdaSpace{}%
\AgdaBound{A}\AgdaSpace{}%
\AgdaSymbol{→}\AgdaSpace{}%
\AgdaDatatype{Term}\AgdaSpace{}%
\AgdaBound{𝔽}\AgdaSpace{}%
\AgdaBound{A}\<%
\\
\>[0]\AgdaFunction{return}\AgdaSpace{}%
\AgdaSymbol{=}\AgdaSpace{}%
\AgdaInductiveConstructor{var}\<%
\\
\\[\AgdaEmptyExtraSkip]%
\>[0]\AgdaOperator{\AgdaFunction{\AgdaUnderscore{}>>=\AgdaUnderscore{}}}\AgdaSpace{}%
\AgdaSymbol{:}\AgdaSpace{}%
\AgdaSymbol{∀}\AgdaSpace{}%
\AgdaSymbol{\{}\AgdaBound{𝔽}\AgdaSymbol{\}}\AgdaSpace{}%
\AgdaSymbol{\{}\AgdaBound{A}\AgdaSpace{}%
\AgdaBound{B}\AgdaSymbol{\}}\AgdaSpace{}%
\AgdaSymbol{→}\AgdaSpace{}%
\AgdaDatatype{Term}\AgdaSpace{}%
\AgdaBound{𝔽}\AgdaSpace{}%
\AgdaBound{A}\AgdaSpace{}%
\AgdaSymbol{→}\AgdaSpace{}%
\AgdaSymbol{(}\AgdaBound{A}\AgdaSpace{}%
\AgdaSymbol{→}\AgdaSpace{}%
\AgdaDatatype{Term}\AgdaSpace{}%
\AgdaBound{𝔽}\AgdaSpace{}%
\AgdaBound{B}\AgdaSymbol{)}\AgdaSpace{}%
\AgdaSymbol{→}\AgdaSpace{}%
\AgdaDatatype{Term}\AgdaSpace{}%
\AgdaBound{𝔽}\AgdaSpace{}%
\AgdaBound{B}\<%
\\
\>[0]\AgdaInductiveConstructor{var}\AgdaSpace{}%
\AgdaBound{x}%
\>[12]\AgdaOperator{\AgdaFunction{>>=}}\AgdaSpace{}%
\AgdaBound{f}\AgdaSpace{}%
\AgdaSymbol{=}%
\>[21]\AgdaBound{f}\AgdaSpace{}%
\AgdaBound{x}\<%
\\
\>[0]\AgdaInductiveConstructor{op}\AgdaSpace{}%
\AgdaSymbol{(}\AgdaBound{o}\AgdaSpace{}%
\AgdaOperator{\AgdaInductiveConstructor{,}}\AgdaSpace{}%
\AgdaBound{k}\AgdaSymbol{)}%
\>[12]\AgdaOperator{\AgdaFunction{>>=}}\AgdaSpace{}%
\AgdaBound{f}\AgdaSpace{}%
\AgdaSymbol{=}%
\>[21]\AgdaInductiveConstructor{op}\AgdaSpace{}%
\AgdaSymbol{(}\AgdaBound{o}\AgdaSpace{}%
\AgdaOperator{\AgdaInductiveConstructor{,}}\AgdaSpace{}%
\AgdaOperator{\AgdaFunction{\AgdaUnderscore{}>>=}}\AgdaSpace{}%
\AgdaBound{f}\AgdaSpace{}%
\AgdaOperator{\AgdaFunction{∘}}\AgdaSpace{}%
\AgdaBound{k}\AgdaSymbol{)}\<%
\end{code}\end{center}
We also provide a helper function \AgdaFunction{perform}, which constructs a \AgdaDatatype{Term} that performs an operation and immediately returns its result:
\begin{center}\begin{code}%
\>[0]\AgdaFunction{perform}\AgdaSpace{}%
\AgdaSymbol{:}\AgdaSpace{}%
\AgdaSymbol{∀}\AgdaSpace{}%
\AgdaSymbol{\{}\AgdaBound{𝔽}\AgdaSymbol{\}}\AgdaSpace{}%
\AgdaSymbol{(}\AgdaBound{o}\AgdaSpace{}%
\AgdaSymbol{:}\AgdaSpace{}%
\AgdaField{Op}\AgdaSpace{}%
\AgdaBound{𝔽}\AgdaSymbol{)}\AgdaSpace{}%
\AgdaSymbol{→}\AgdaSpace{}%
\AgdaDatatype{Term}\AgdaSpace{}%
\AgdaBound{𝔽}\AgdaSpace{}%
\AgdaSymbol{(}\AgdaField{Arity}\AgdaSpace{}%
\AgdaBound{𝔽}\AgdaSpace{}%
\AgdaBound{o}\AgdaSymbol{)}\<%
\\
\>[0]\AgdaFunction{perform}\AgdaSpace{}%
\AgdaBound{o}\AgdaSpace{}%
\AgdaSymbol{=}\AgdaSpace{}%
\AgdaInductiveConstructor{op}\AgdaSpace{}%
\AgdaSymbol{(}\AgdaBound{o}\AgdaSpace{}%
\AgdaOperator{\AgdaInductiveConstructor{,}}\AgdaSpace{}%
\AgdaInductiveConstructor{var}\AgdaSymbol{)}\<%
\end{code}\end{center}

\begin{code}[hide]%
\>[0]\AgdaKeyword{instance}\<%
\\
\>[0][@{}l@{\AgdaIndent{0}}]%
\>[2]\AgdaFunction{term-monad}\AgdaSpace{}%
\AgdaSymbol{:}\AgdaSpace{}%
\AgdaSymbol{∀}\AgdaSpace{}%
\AgdaSymbol{\{}\AgdaBound{𝔽}\AgdaSymbol{\}}\AgdaSpace{}%
\AgdaSymbol{→}\AgdaSpace{}%
\AgdaRecord{RawMonad}\AgdaSpace{}%
\AgdaSymbol{(}\AgdaDatatype{Term}\AgdaSpace{}%
\AgdaBound{𝔽}\AgdaSymbol{)}\<%
\\
\>[2]\AgdaFunction{term-monad}\AgdaSpace{}%
\AgdaSymbol{=}\AgdaSpace{}%
\AgdaFunction{mkRawMonad}\AgdaSpace{}%
\AgdaSymbol{\AgdaUnderscore{}}\AgdaSpace{}%
\AgdaFunction{return}\AgdaSpace{}%
\AgdaOperator{\AgdaFunction{\AgdaUnderscore{}>>=\AgdaUnderscore{}}}\<%
\end{code}

\subsection{Effect Handlers}

One of the reasons why \AgdaDatatype{Term}s are called the free algebra of $\mathbb{F}$ is that they are freely interpretable.
Given another $\mathbb{F}$-algebra $X$ and a substitution of variables from $X$, we can interpret a term as an element of $X$:
\begin{center}\begin{code}%
\>[0]\AgdaFunction{interp}\AgdaSpace{}%
\AgdaSymbol{:}\AgdaSpace{}%
\AgdaSymbol{∀}\AgdaSpace{}%
\AgdaSymbol{\{}\AgdaBound{𝔽}\AgdaSymbol{\}}\AgdaSpace{}%
\AgdaSymbol{\{}\AgdaBound{X}\AgdaSpace{}%
\AgdaBound{A}\AgdaSymbol{\}}\AgdaSpace{}%
\AgdaSymbol{→}\AgdaSpace{}%
\AgdaBound{𝔽}\AgdaSpace{}%
\AgdaOperator{\AgdaFunction{-Alg[}}\AgdaSpace{}%
\AgdaBound{X}\AgdaSpace{}%
\AgdaOperator{\AgdaFunction{]}}\AgdaSpace{}%
\AgdaSymbol{→}\AgdaSpace{}%
\AgdaSymbol{(}\AgdaBound{A}\AgdaSpace{}%
\AgdaSymbol{→}\AgdaSpace{}%
\AgdaBound{X}\AgdaSymbol{)}\AgdaSpace{}%
\AgdaSymbol{→}\AgdaSpace{}%
\AgdaDatatype{Term}\AgdaSpace{}%
\AgdaBound{𝔽}\AgdaSpace{}%
\AgdaBound{A}\AgdaSpace{}%
\AgdaSymbol{→}\AgdaSpace{}%
\AgdaBound{X}\<%
\\
\>[0]\AgdaFunction{interp}\AgdaSpace{}%
\AgdaBound{alg}\AgdaSpace{}%
\AgdaBound{f}\AgdaSpace{}%
\AgdaSymbol{(}\AgdaInductiveConstructor{var}\AgdaSpace{}%
\AgdaBound{x}\AgdaSymbol{)}%
\>[27]\AgdaSymbol{=}\AgdaSpace{}%
\AgdaBound{f}\AgdaSpace{}%
\AgdaBound{x}\<%
\\
\>[0]\AgdaFunction{interp}\AgdaSpace{}%
\AgdaBound{alg}\AgdaSpace{}%
\AgdaBound{f}\AgdaSpace{}%
\AgdaSymbol{(}\AgdaInductiveConstructor{op}\AgdaSpace{}%
\AgdaSymbol{(}\AgdaBound{o}\AgdaSpace{}%
\AgdaOperator{\AgdaInductiveConstructor{,}}\AgdaSpace{}%
\AgdaBound{k}\AgdaSymbol{))}%
\>[27]\AgdaSymbol{=}\AgdaSpace{}%
\AgdaBound{alg}\AgdaSpace{}%
\AgdaSymbol{(}\AgdaBound{o}\AgdaSpace{}%
\AgdaOperator{\AgdaInductiveConstructor{,}}\AgdaSpace{}%
\AgdaFunction{interp}\AgdaSpace{}%
\AgdaBound{alg}\AgdaSpace{}%
\AgdaBound{f}\AgdaSpace{}%
\AgdaOperator{\AgdaFunction{∘}}\AgdaSpace{}%
\AgdaBound{k}\AgdaSymbol{)}\<%
\end{code}\end{center}
For variables, \AgdaFunction{interp} uses the substitution $f$ to map them to $X$.
For operations, \AgdaFunction{interp} first recursively interprets the arguments and then uses ${alg}$ to make a new element of $X$ from a fully applied operation.

In programming, \AgdaFunction{interp} forms the foundation of effect handlers.
It lets us systematically handle uninterpreted effects of a computation.

\section{Choreographic Programming with Algebraic Effects}
\label{sec:cp-alg-eff}

In this section, we use the algebraic effects framework from the previous section to implement a CP language.
Due to lack of space, we only consider a minimum CP language with only local computations and communication, omitting features like conditionals and recursion for now.
We start by defining processes~(\Cref{sec:process}), which are the results of endpoint projection.
Then, we move on to defining choreographies~(\Cref{sec:choreo}).
Unlike previous library-level CP languages, our choreographies abstract over a particular representation of located values, allowing us to erase them and avoid non-totality.
Finally, we define endpoint projection as location-specific effect handlers for choreographies~(\Cref{sec:epp}).

We assume a local language of signature 𝕃 that each node uses for local computations, and we parameterize our CP language by it.
We use locations Loc to refer to nodes in a distributed system and define them as Strings.
However, any data type with decidable equality would suffice.

\begin{code}[hide]%
\>[0]\AgdaKeyword{open}\AgdaSpace{}%
\AgdaKeyword{import}\AgdaSpace{}%
\AgdaModule{02-algeff}\AgdaSpace{}%
\AgdaKeyword{hiding}\AgdaSpace{}%
\AgdaSymbol{(}\AgdaFunction{return}\AgdaSymbol{;}\AgdaSpace{}%
\AgdaOperator{\AgdaFunction{\AgdaUnderscore{}>>=\AgdaUnderscore{}}}\AgdaSymbol{)}\<%
\\
\\[\AgdaEmptyExtraSkip]%
\>[0]\AgdaKeyword{module}\AgdaSpace{}%
\AgdaModule{03-choreo}\AgdaSpace{}%
\AgdaSymbol{(}\AgdaBound{𝕃}\AgdaSpace{}%
\AgdaSymbol{:}\AgdaSpace{}%
\AgdaRecord{Sig}\AgdaSymbol{)}\AgdaSpace{}%
\AgdaKeyword{where}\<%
\\
\\[\AgdaEmptyExtraSkip]%
\>[0]\AgdaKeyword{open}\AgdaSpace{}%
\AgdaKeyword{import}\AgdaSpace{}%
\AgdaModule{Data.Product}\AgdaSpace{}%
\AgdaKeyword{using}\AgdaSpace{}%
\AgdaSymbol{(}\AgdaOperator{\AgdaInductiveConstructor{\AgdaUnderscore{},\AgdaUnderscore{}}}\AgdaSymbol{)}\<%
\\
\>[0]\AgdaKeyword{open}\AgdaSpace{}%
\AgdaKeyword{import}\AgdaSpace{}%
\AgdaModule{Data.String}\AgdaSpace{}%
\AgdaSymbol{as}\AgdaSpace{}%
\AgdaModule{String}\AgdaSpace{}%
\AgdaKeyword{using}\AgdaSymbol{(}\AgdaPostulate{String}\AgdaSymbol{)}\<%
\\
\>[0]\AgdaKeyword{open}\AgdaSpace{}%
\AgdaKeyword{import}\AgdaSpace{}%
\AgdaModule{Effect.Monad}\AgdaSpace{}%
\AgdaKeyword{using}\AgdaSpace{}%
\AgdaSymbol{(}\AgdaRecord{RawMonad}\AgdaSymbol{)}\<%
\\
\>[0]\AgdaKeyword{open}\AgdaSpace{}%
\AgdaKeyword{import}\AgdaSpace{}%
\AgdaModule{Level}\AgdaSpace{}%
\AgdaKeyword{using}\AgdaSpace{}%
\AgdaSymbol{(}\AgdaPostulate{Level}\AgdaSymbol{;}\AgdaSpace{}%
\AgdaPrimitive{Setω}\AgdaSymbol{)}\<%
\\
\>[0]\AgdaKeyword{open}\AgdaSpace{}%
\AgdaKeyword{import}\AgdaSpace{}%
\AgdaModule{Relation.Binary.PropositionalEquality}\AgdaSpace{}%
\AgdaKeyword{using}\AgdaSpace{}%
\AgdaSymbol{(}\AgdaOperator{\AgdaDatatype{\AgdaUnderscore{}≡\AgdaUnderscore{}}}\AgdaSymbol{)}\<%
\\
\>[0]\AgdaKeyword{open}\AgdaSpace{}%
\AgdaKeyword{import}\AgdaSpace{}%
\AgdaModule{Relation.Nullary}\AgdaSpace{}%
\AgdaKeyword{using}\AgdaSpace{}%
\AgdaSymbol{(}\AgdaRecord{Dec}\AgdaSymbol{;}\AgdaSpace{}%
\AgdaInductiveConstructor{yes}\AgdaSymbol{;}\AgdaSpace{}%
\AgdaInductiveConstructor{no}\AgdaSymbol{)}\<%
\\
\\[\AgdaEmptyExtraSkip]%
\>[0]\AgdaKeyword{open}\AgdaSpace{}%
\AgdaModule{RawMonad}\AgdaSpace{}%
\AgdaSymbol{⦃...⦄}\<%
\\
\\[\AgdaEmptyExtraSkip]%
\>[0]\AgdaKeyword{infix}\AgdaSpace{}%
\AgdaNumber{20}\AgdaSpace{}%
\AgdaOperator{\AgdaFunction{\AgdaUnderscore{}▷\AgdaUnderscore{}}}\<%
\\
\>[0]\AgdaKeyword{infix}\AgdaSpace{}%
\AgdaNumber{20}\AgdaSpace{}%
\AgdaOperator{\AgdaFunction{\AgdaUnderscore{}⇒\AgdaUnderscore{}◇\AgdaUnderscore{}}}\<%
\\
\>[0]\AgdaFunction{Loc}\AgdaSpace{}%
\AgdaSymbol{:}\AgdaSpace{}%
\AgdaPrimitive{Set}\<%
\\
\>[0]\AgdaFunction{Loc}\AgdaSpace{}%
\AgdaSymbol{=}\AgdaSpace{}%
\AgdaPostulate{String}\<%
\\
\\[\AgdaEmptyExtraSkip]%
\>[0]\AgdaComment{--\ the\ opaque\ prevents\ Agda\ from\ unfolding\ the\ definition,\ making\ the\ code\ more\ readable\ in\ certain\ cases}\<%
\\
\>[0]\AgdaKeyword{opaque}\<%
\\
\>[0][@{}l@{\AgdaIndent{0}}]%
\>[2]\AgdaOperator{\AgdaFunction{\AgdaUnderscore{}≟\AgdaUnderscore{}}}\AgdaSpace{}%
\AgdaSymbol{:}\AgdaSpace{}%
\AgdaSymbol{(}\AgdaBound{l}\AgdaSpace{}%
\AgdaBound{l′}\AgdaSpace{}%
\AgdaSymbol{:}\AgdaSpace{}%
\AgdaFunction{Loc}\AgdaSymbol{)}\AgdaSpace{}%
\AgdaSymbol{→}\AgdaSpace{}%
\AgdaRecord{Dec}\AgdaSpace{}%
\AgdaSymbol{(}\AgdaBound{l}\AgdaSpace{}%
\AgdaOperator{\AgdaDatatype{≡}}\AgdaSpace{}%
\AgdaBound{l′}\AgdaSymbol{)}\<%
\\
\>[2]\AgdaOperator{\AgdaFunction{\AgdaUnderscore{}≟\AgdaUnderscore{}}}\AgdaSpace{}%
\AgdaSymbol{=}\AgdaSpace{}%
\AgdaOperator{\AgdaFunction{String.\AgdaUnderscore{}≟\AgdaUnderscore{}}}\<%
\\
\\[\AgdaEmptyExtraSkip]%
\>[0]\AgdaComment{--\ the\ stdlib's\ ⊤\ is\ not\ universe-polymorphic}\<%
\\
\>[0]\AgdaKeyword{record}\AgdaSpace{}%
\AgdaRecord{⊤}\AgdaSpace{}%
\AgdaSymbol{\{}\AgdaBound{ℓ}\AgdaSpace{}%
\AgdaSymbol{:}\AgdaSpace{}%
\AgdaPostulate{Level}\AgdaSymbol{\}}\AgdaSpace{}%
\AgdaSymbol{:}\AgdaSpace{}%
\AgdaPrimitive{Set}\AgdaSpace{}%
\AgdaBound{ℓ}\AgdaSpace{}%
\AgdaKeyword{where}\<%
\\
\>[0][@{}l@{\AgdaIndent{0}}]%
\>[2]\AgdaKeyword{constructor}\AgdaSpace{}%
\AgdaInductiveConstructor{tt}\<%
\end{code}

\subsection{Processes}
\label{sec:process}

\begin{code}[hide]%
\>[0]\AgdaComment{--\ a\ seperate\ modules\ avoid\ name\ conflicts}\<%
\\
\>[0]\AgdaKeyword{module}\AgdaSpace{}%
\AgdaModule{Process}\AgdaSpace{}%
\AgdaKeyword{where}\<%
\end{code}

\begin{figure}[ht]

\begin{minipage}{.45\textwidth}
  \begin{center}\begin{code}%
\>[0][@{}l@{\AgdaIndent{1}}]%
\>[2]\AgdaKeyword{data}\AgdaSpace{}%
\AgdaDatatype{Op}\AgdaSpace{}%
\AgdaSymbol{:}\AgdaSpace{}%
\AgdaPrimitive{Set₁}\AgdaSpace{}%
\AgdaKeyword{where}\<%
\\
\>[2][@{}l@{\AgdaIndent{0}}]%
\>[4]\AgdaInductiveConstructor{`locally}\AgdaSpace{}%
\AgdaSymbol{:}\AgdaSpace{}%
\AgdaSymbol{∀}\AgdaSpace{}%
\AgdaSymbol{\{}\AgdaBound{A}\AgdaSymbol{\}}\AgdaSpace{}%
\AgdaSymbol{→}\AgdaSpace{}%
\AgdaDatatype{Term}\AgdaSpace{}%
\AgdaBound{𝕃}\AgdaSpace{}%
\AgdaBound{A}\AgdaSpace{}%
\AgdaSymbol{→}\AgdaSpace{}%
\AgdaDatatype{Op}\<%
\\
\>[4]\AgdaInductiveConstructor{`send}%
\>[13]\AgdaSymbol{:}\AgdaSpace{}%
\AgdaSymbol{∀}\AgdaSpace{}%
\AgdaSymbol{\{}\AgdaBound{A}\AgdaSpace{}%
\AgdaSymbol{:}\AgdaSpace{}%
\AgdaPrimitive{Set}\AgdaSymbol{\}}\AgdaSpace{}%
\AgdaSymbol{→}\AgdaSpace{}%
\AgdaFunction{Loc}\AgdaSpace{}%
\AgdaSymbol{→}\AgdaSpace{}%
\AgdaBound{A}\AgdaSpace{}%
\AgdaSymbol{→}\AgdaSpace{}%
\AgdaDatatype{Op}\<%
\\
\>[4]\AgdaInductiveConstructor{`recv}%
\>[13]\AgdaSymbol{:}\AgdaSpace{}%
\AgdaSymbol{∀}\AgdaSpace{}%
\AgdaSymbol{\{}\AgdaBound{A}\AgdaSpace{}%
\AgdaSymbol{:}\AgdaSpace{}%
\AgdaPrimitive{Set}\AgdaSymbol{\}}\AgdaSpace{}%
\AgdaSymbol{→}\AgdaSpace{}%
\AgdaFunction{Loc}\AgdaSpace{}%
\AgdaSymbol{→}\AgdaSpace{}%
\AgdaDatatype{Op}\<%
\\
\\[\AgdaEmptyExtraSkip]%
\>[2]\AgdaFunction{Arity}\AgdaSpace{}%
\AgdaSymbol{:}\AgdaSpace{}%
\AgdaDatatype{Op}\AgdaSpace{}%
\AgdaSymbol{→}\AgdaSpace{}%
\AgdaPrimitive{Set}\<%
\\
\>[2]\AgdaFunction{Arity}\AgdaSpace{}%
\AgdaSymbol{(}\AgdaInductiveConstructor{`locally}\AgdaSpace{}%
\AgdaSymbol{\{}\AgdaBound{A}\AgdaSymbol{\}}\AgdaSpace{}%
\AgdaSymbol{\AgdaUnderscore{})}\AgdaSpace{}%
\AgdaSymbol{=}\AgdaSpace{}%
\AgdaBound{A}\<%
\\
\>[2]\AgdaFunction{Arity}\AgdaSpace{}%
\AgdaSymbol{(}\AgdaInductiveConstructor{`send}\AgdaSpace{}%
\AgdaSymbol{\AgdaUnderscore{}}\AgdaSpace{}%
\AgdaSymbol{\AgdaUnderscore{})}\AgdaSpace{}%
\AgdaSymbol{=}\AgdaSpace{}%
\AgdaRecord{⊤}\<%
\\
\>[2]\AgdaFunction{Arity}\AgdaSpace{}%
\AgdaSymbol{(}\AgdaInductiveConstructor{`recv}\AgdaSpace{}%
\AgdaSymbol{\{}\AgdaBound{A}\AgdaSymbol{\}}\AgdaSpace{}%
\AgdaSymbol{\AgdaUnderscore{})}\AgdaSpace{}%
\AgdaSymbol{=}\AgdaSpace{}%
\AgdaBound{A}\<%
\\
\\[\AgdaEmptyExtraSkip]%
\>[2]\AgdaFunction{ℙ}\AgdaSpace{}%
\AgdaSymbol{:}\AgdaSpace{}%
\AgdaRecord{Sig}\<%
\\
\>[2]\AgdaFunction{ℙ}\AgdaSpace{}%
\AgdaSymbol{=}\AgdaSpace{}%
\AgdaDatatype{Op}\AgdaSpace{}%
\AgdaOperator{\AgdaInductiveConstructor{◁}}\AgdaSpace{}%
\AgdaFunction{Arity}\<%
\end{code}\end{center}
\end{minipage}
\hfill\vline\hfill
\begin{minipage}{.45\textwidth}
  \begin{center}\begin{code}%
\>[2]\AgdaFunction{Process}\AgdaSpace{}%
\AgdaSymbol{:}\AgdaSpace{}%
\AgdaPrimitive{Set}\AgdaSpace{}%
\AgdaSymbol{→}\AgdaSpace{}%
\AgdaPrimitive{Set₁}\<%
\\
\>[2]\AgdaFunction{Process}\AgdaSpace{}%
\AgdaBound{A}\AgdaSpace{}%
\AgdaSymbol{=}\AgdaSpace{}%
\AgdaDatatype{Term}\AgdaSpace{}%
\AgdaFunction{ℙ}\AgdaSpace{}%
\AgdaBound{A}\<%
\\
\\[\AgdaEmptyExtraSkip]%
\>[2]\AgdaFunction{locally}\AgdaSpace{}%
\AgdaSymbol{:}\AgdaSpace{}%
\AgdaSymbol{∀}\AgdaSpace{}%
\AgdaSymbol{\{}\AgdaBound{A}\AgdaSymbol{\}}\AgdaSpace{}%
\AgdaSymbol{→}\AgdaSpace{}%
\AgdaDatatype{Term}\AgdaSpace{}%
\AgdaBound{𝕃}\AgdaSpace{}%
\AgdaBound{A}\AgdaSpace{}%
\AgdaSymbol{→}\AgdaSpace{}%
\AgdaFunction{Process}\AgdaSpace{}%
\AgdaBound{A}\<%
\\
\>[2]\AgdaFunction{locally}\AgdaSpace{}%
\AgdaBound{t}\AgdaSpace{}%
\AgdaSymbol{=}\AgdaSpace{}%
\AgdaFunction{perform}\AgdaSpace{}%
\AgdaSymbol{(}\AgdaInductiveConstructor{`locally}\AgdaSpace{}%
\AgdaBound{t}\AgdaSymbol{)}\<%
\\
\\[\AgdaEmptyExtraSkip]%
\>[2]\AgdaFunction{send}\AgdaSpace{}%
\AgdaSymbol{:}\AgdaSpace{}%
\AgdaSymbol{∀}\AgdaSpace{}%
\AgdaSymbol{\{}\AgdaBound{A}\AgdaSymbol{\}}\AgdaSpace{}%
\AgdaSymbol{→}\AgdaSpace{}%
\AgdaFunction{Loc}\AgdaSpace{}%
\AgdaSymbol{→}\AgdaSpace{}%
\AgdaBound{A}\AgdaSpace{}%
\AgdaSymbol{→}\AgdaSpace{}%
\AgdaFunction{Process}\AgdaSpace{}%
\AgdaRecord{⊤}\<%
\\
\>[2]\AgdaFunction{send}\AgdaSpace{}%
\AgdaBound{l}\AgdaSpace{}%
\AgdaBound{a}\AgdaSpace{}%
\AgdaSymbol{=}\AgdaSpace{}%
\AgdaFunction{perform}\AgdaSpace{}%
\AgdaSymbol{(}\AgdaInductiveConstructor{`send}\AgdaSpace{}%
\AgdaBound{l}\AgdaSpace{}%
\AgdaBound{a}\AgdaSymbol{)}\<%
\\
\\[\AgdaEmptyExtraSkip]%
\>[2]\AgdaFunction{recv}\AgdaSpace{}%
\AgdaSymbol{:}\AgdaSpace{}%
\AgdaSymbol{∀}\AgdaSpace{}%
\AgdaSymbol{\{}\AgdaBound{A}\AgdaSymbol{\}}\AgdaSpace{}%
\AgdaSymbol{→}\AgdaSpace{}%
\AgdaFunction{Loc}\AgdaSpace{}%
\AgdaSymbol{→}\AgdaSpace{}%
\AgdaFunction{Process}\AgdaSpace{}%
\AgdaBound{A}\<%
\\
\>[2]\AgdaFunction{recv}\AgdaSpace{}%
\AgdaSymbol{\{}\AgdaBound{A}\AgdaSymbol{\}}\AgdaSpace{}%
\AgdaBound{l}\AgdaSpace{}%
\AgdaSymbol{=}\AgdaSpace{}%
\AgdaFunction{perform}\AgdaSpace{}%
\AgdaSymbol{(}\AgdaInductiveConstructor{`recv}\AgdaSpace{}%
\AgdaSymbol{\{}\AgdaBound{A}\AgdaSymbol{\}}\AgdaSpace{}%
\AgdaBound{l}\AgdaSymbol{)}\<%
\end{code}\end{center}
\end{minipage}

\caption{Processes as Algebraic Effects}
\label{fig:process}
\end{figure}

\Cref{fig:process} presents processes as algebraic effects.
Signature ℙ specifies the three operations of processes and their arity:
\begin{itemize}
\item
  \AgdaInductiveConstructor{`locally} performs a local computation of type \AgdaDatatype{Term} 𝕃 $A$ and returns a value of type $A$.
\item
  \AgdaInductiveConstructor{`send} sends a message of type $A$ to a location and returns a unit value.
\item
  \AgdaInductiveConstructor{`recv} receives a message from a location and returns a value of type $A$.
  Here, the performer of the \AgdaInductiveConstructor{`recv} needs to specify what type of value it is expected to receive.
\end{itemize}
We also define \AgdaDatatype{Process} as a shorthand for terms using operations from ℙ. Finally, we define the helper functions \AgdaFunction{send}, \AgdaFunction{receive}, and \AgdaFunction{locally}.

\subsection{Choreographies}
\label{sec:choreo}

\begin{figure}[ht]

\begin{minipage}{.45\textwidth}
  \begin{center}\begin{code}%
\>[0]\AgdaFunction{At}\AgdaSpace{}%
\AgdaSymbol{:}\AgdaSpace{}%
\AgdaPrimitive{Setω}\<%
\\
\>[0]\AgdaFunction{At}\AgdaSpace{}%
\AgdaSymbol{=}\AgdaSpace{}%
\AgdaSymbol{∀}\AgdaSpace{}%
\AgdaSymbol{\{}\AgdaBound{ℓ}\AgdaSymbol{\}}\AgdaSpace{}%
\AgdaSymbol{→}\AgdaSpace{}%
\AgdaPrimitive{Set}\AgdaSpace{}%
\AgdaBound{ℓ}\AgdaSpace{}%
\AgdaSymbol{→}\AgdaSpace{}%
\AgdaFunction{Loc}\AgdaSpace{}%
\AgdaSymbol{→}\AgdaSpace{}%
\AgdaPrimitive{Set}\AgdaSpace{}%
\AgdaBound{ℓ}\<%
\\
\\[\AgdaEmptyExtraSkip]%
\>[0]\AgdaFunction{focus}\AgdaSpace{}%
\AgdaSymbol{:}\AgdaSpace{}%
\AgdaFunction{Loc}\AgdaSpace{}%
\AgdaSymbol{→}\AgdaSpace{}%
\AgdaFunction{At}\<%
\\
\>[0]\AgdaFunction{focus}\AgdaSpace{}%
\AgdaBound{l}\AgdaSpace{}%
\AgdaBound{A}\AgdaSpace{}%
\AgdaBound{s}\AgdaSpace{}%
\AgdaKeyword{with}\AgdaSpace{}%
\AgdaBound{l}\AgdaSpace{}%
\AgdaOperator{\AgdaFunction{≟}}\AgdaSpace{}%
\AgdaBound{s}\<%
\\
\>[0]\AgdaSymbol{...}\AgdaSpace{}%
\AgdaSymbol{|}\AgdaSpace{}%
\AgdaInductiveConstructor{yes}\AgdaSpace{}%
\AgdaSymbol{\AgdaUnderscore{}}%
\>[13]\AgdaSymbol{=}\AgdaSpace{}%
\AgdaBound{A}\<%
\\
\>[0]\AgdaSymbol{...}\AgdaSpace{}%
\AgdaSymbol{|}\AgdaSpace{}%
\AgdaInductiveConstructor{no}%
\>[10]\AgdaSymbol{\AgdaUnderscore{}}%
\>[13]\AgdaSymbol{=}\AgdaSpace{}%
\AgdaRecord{⊤}\<%
\\
\\[\AgdaEmptyExtraSkip]%
\>[0]\AgdaKeyword{module}\AgdaSpace{}%
\AgdaModule{\AgdaUnderscore{}}\AgdaSpace{}%
\AgdaSymbol{(}\AgdaOperator{\AgdaBound{\AgdaUnderscore{}＠\AgdaUnderscore{}}}\AgdaSpace{}%
\AgdaSymbol{:}\AgdaSpace{}%
\AgdaFunction{At}\AgdaSymbol{)}\AgdaSpace{}%
\AgdaKeyword{where}\<%
\\
\\[\AgdaEmptyExtraSkip]%
\>[0][@{}l@{\AgdaIndent{0}}]%
\>[2]\AgdaKeyword{data}\AgdaSpace{}%
\AgdaDatatype{Op}\AgdaSpace{}%
\AgdaSymbol{:}\AgdaSpace{}%
\AgdaPrimitive{Set₁}\AgdaSpace{}%
\AgdaKeyword{where}\<%
\\
\>[2][@{}l@{\AgdaIndent{0}}]%
\>[4]\AgdaInductiveConstructor{`comm}\AgdaSpace{}%
\AgdaSymbol{:}%
\>[13]\AgdaSymbol{∀}\AgdaSpace{}%
\AgdaSymbol{\{}\AgdaBound{A}\AgdaSymbol{\}}\AgdaSpace{}%
\AgdaSymbol{(}\AgdaBound{s}\AgdaSpace{}%
\AgdaBound{r}\AgdaSpace{}%
\AgdaSymbol{:}\AgdaSpace{}%
\AgdaFunction{Loc}\AgdaSymbol{)}\AgdaSpace{}%
\AgdaSymbol{→}\<%
\\
\>[13]\AgdaSymbol{(}\AgdaDatatype{Term}\AgdaSpace{}%
\AgdaBound{𝕃}\AgdaSpace{}%
\AgdaBound{A}\AgdaSymbol{)}\AgdaSpace{}%
\AgdaOperator{\AgdaBound{＠}}\AgdaSpace{}%
\AgdaBound{s}\AgdaSpace{}%
\AgdaSymbol{→}\AgdaSpace{}%
\AgdaDatatype{Op}\<%
\\
\\[\AgdaEmptyExtraSkip]%
\>[2]\AgdaFunction{Arity}\AgdaSpace{}%
\AgdaSymbol{:}\AgdaSpace{}%
\AgdaDatatype{Op}\AgdaSpace{}%
\AgdaSymbol{→}\AgdaSpace{}%
\AgdaPrimitive{Set}\AgdaSpace{}%
\AgdaSymbol{\AgdaUnderscore{}}\<%
\\
\>[2]\AgdaFunction{Arity}\AgdaSpace{}%
\AgdaSymbol{(}\AgdaInductiveConstructor{`comm}\AgdaSpace{}%
\AgdaSymbol{\{}\AgdaBound{A}\AgdaSymbol{\}}\AgdaSpace{}%
\AgdaSymbol{\AgdaUnderscore{}}\AgdaSpace{}%
\AgdaBound{r}\AgdaSpace{}%
\AgdaSymbol{\AgdaUnderscore{})}\AgdaSpace{}%
\AgdaSymbol{=}\AgdaSpace{}%
\AgdaBound{A}\AgdaSpace{}%
\AgdaOperator{\AgdaBound{＠}}\AgdaSpace{}%
\AgdaBound{r}\<%
\\
\\[\AgdaEmptyExtraSkip]%
\>[2]\AgdaFunction{ℂ}\AgdaSpace{}%
\AgdaSymbol{:}\AgdaSpace{}%
\AgdaRecord{Sig}\<%
\\
\>[2]\AgdaFunction{ℂ}\AgdaSpace{}%
\AgdaSymbol{=}\AgdaSpace{}%
\AgdaDatatype{Op}\AgdaSpace{}%
\AgdaOperator{\AgdaInductiveConstructor{◁}}\AgdaSpace{}%
\AgdaFunction{Arity}\<%
\\
\>[2]\AgdaComment{--\ the\ module\ ends\ here}\<%
\end{code}\end{center}
\end{minipage}
\hfill\vline\hfill
\begin{minipage}{.45\textwidth}
  \begin{center}\begin{code}%
\>[0]\AgdaFunction{Choreo}\AgdaSpace{}%
\AgdaSymbol{:}\AgdaSpace{}%
\AgdaSymbol{(}\AgdaFunction{At}\AgdaSpace{}%
\AgdaSymbol{→}\AgdaSpace{}%
\AgdaPrimitive{Set}\AgdaSymbol{)}\AgdaSpace{}%
\AgdaSymbol{→}\AgdaSpace{}%
\AgdaPrimitive{Setω}\<%
\\
\>[0]\AgdaFunction{Choreo}\AgdaSpace{}%
\AgdaBound{F}\AgdaSpace{}%
\AgdaSymbol{=}\<%
\\
\>[0][@{}l@{\AgdaIndent{0}}]%
\>[2]\AgdaSymbol{∀}\AgdaSpace{}%
\AgdaSymbol{\{}\AgdaOperator{\AgdaBound{\AgdaUnderscore{}＠\AgdaUnderscore{}}}\AgdaSpace{}%
\AgdaSymbol{:}\AgdaSpace{}%
\AgdaFunction{At}\AgdaSymbol{\}}\<%
\\
\>[2]\AgdaSymbol{\{\{}\AgdaBound{\AgdaUnderscore{}}\AgdaSpace{}%
\AgdaSymbol{:}\AgdaSpace{}%
\AgdaSymbol{∀}\AgdaSpace{}%
\AgdaSymbol{\{}\AgdaBound{ℓ}\AgdaSymbol{\}}\AgdaSpace{}%
\AgdaSymbol{\{}\AgdaBound{l}\AgdaSymbol{\}}\AgdaSpace{}%
\AgdaSymbol{→}\AgdaSpace{}%
\AgdaRecord{RawMonad}\AgdaSpace{}%
\AgdaSymbol{\{}\AgdaBound{ℓ}\AgdaSymbol{\}}\AgdaSpace{}%
\AgdaSymbol{(}\AgdaOperator{\AgdaBound{\AgdaUnderscore{}＠}}\AgdaSpace{}%
\AgdaBound{l}\AgdaSymbol{)\}\}}\AgdaSpace{}%
\AgdaSymbol{→}\<%
\\
\>[2]\AgdaDatatype{Term}\AgdaSpace{}%
\AgdaSymbol{(}\AgdaFunction{ℂ}\AgdaSpace{}%
\AgdaOperator{\AgdaBound{\AgdaUnderscore{}＠\AgdaUnderscore{}}}\AgdaSymbol{)}\AgdaSpace{}%
\AgdaSymbol{(}\AgdaBound{F}\AgdaSpace{}%
\AgdaOperator{\AgdaBound{\AgdaUnderscore{}＠\AgdaUnderscore{}}}\AgdaSymbol{)}\<%
\\
\\[\AgdaEmptyExtraSkip]%
\>[0]\AgdaOperator{\AgdaFunction{\AgdaUnderscore{}▷\AgdaUnderscore{}}}\AgdaSpace{}%
\AgdaSymbol{:}%
\>[7]\AgdaSymbol{∀}\AgdaSpace{}%
\AgdaSymbol{\{}\AgdaOperator{\AgdaBound{\AgdaUnderscore{}＠\AgdaUnderscore{}}}\AgdaSpace{}%
\AgdaSymbol{:}\AgdaSpace{}%
\AgdaFunction{At}\AgdaSymbol{\}}\AgdaSpace{}%
\AgdaSymbol{\{}\AgdaBound{A}\AgdaSymbol{\}}\AgdaSpace{}%
\AgdaSymbol{→}\<%
\\
\>[7]\AgdaSymbol{(}\AgdaBound{s}\AgdaSpace{}%
\AgdaSymbol{:}\AgdaSpace{}%
\AgdaFunction{Loc}\AgdaSymbol{)}\AgdaSpace{}%
\AgdaSymbol{→}\AgdaSpace{}%
\AgdaSymbol{(}\AgdaDatatype{Term}\AgdaSpace{}%
\AgdaBound{𝕃}\AgdaSpace{}%
\AgdaBound{A}\AgdaSymbol{)}\AgdaSpace{}%
\AgdaOperator{\AgdaBound{＠}}\AgdaSpace{}%
\AgdaBound{s}\AgdaSpace{}%
\AgdaSymbol{→}\<%
\\
\>[7]\AgdaDatatype{Term}\AgdaSpace{}%
\AgdaSymbol{(}\AgdaFunction{ℂ}\AgdaSpace{}%
\AgdaOperator{\AgdaBound{\AgdaUnderscore{}＠\AgdaUnderscore{}}}\AgdaSymbol{)}\AgdaSpace{}%
\AgdaSymbol{(}\AgdaBound{A}\AgdaSpace{}%
\AgdaOperator{\AgdaBound{＠}}\AgdaSpace{}%
\AgdaBound{s}\AgdaSymbol{)}\<%
\\
\>[0]\AgdaBound{s}\AgdaSpace{}%
\AgdaOperator{\AgdaFunction{▷}}\AgdaSpace{}%
\AgdaBound{t}\AgdaSpace{}%
\AgdaSymbol{=}\AgdaSpace{}%
\AgdaFunction{perform}\AgdaSpace{}%
\AgdaSymbol{(}\AgdaInductiveConstructor{`comm}\AgdaSpace{}%
\AgdaBound{s}\AgdaSpace{}%
\AgdaBound{s}\AgdaSpace{}%
\AgdaBound{t}\AgdaSymbol{)}\<%
\\
\\[\AgdaEmptyExtraSkip]%
\>[0]\AgdaOperator{\AgdaFunction{\AgdaUnderscore{}⇒\AgdaUnderscore{}◇\AgdaUnderscore{}}}\AgdaSpace{}%
\AgdaSymbol{:}%
\>[9]\AgdaSymbol{∀}\AgdaSpace{}%
\AgdaSymbol{\{}\AgdaOperator{\AgdaBound{\AgdaUnderscore{}＠\AgdaUnderscore{}}}\AgdaSpace{}%
\AgdaSymbol{:}\AgdaSpace{}%
\AgdaFunction{At}\AgdaSymbol{\}}\AgdaSpace{}%
\AgdaSymbol{\{}\AgdaBound{A}\AgdaSymbol{\}}\AgdaSpace{}%
\AgdaSymbol{→}\<%
\\
\>[9]\AgdaSymbol{(}\AgdaBound{s}\AgdaSpace{}%
\AgdaBound{r}\AgdaSpace{}%
\AgdaSymbol{:}\AgdaSpace{}%
\AgdaFunction{Loc}\AgdaSymbol{)}\AgdaSpace{}%
\AgdaSymbol{→}\AgdaSpace{}%
\AgdaSymbol{(}\AgdaDatatype{Term}\AgdaSpace{}%
\AgdaBound{𝕃}\AgdaSpace{}%
\AgdaBound{A}\AgdaSymbol{)}\AgdaSpace{}%
\AgdaOperator{\AgdaBound{＠}}\AgdaSpace{}%
\AgdaBound{s}\AgdaSpace{}%
\AgdaSymbol{→}\<%
\\
\>[9]\AgdaDatatype{Term}\AgdaSpace{}%
\AgdaSymbol{(}\AgdaFunction{ℂ}\AgdaSpace{}%
\AgdaOperator{\AgdaBound{\AgdaUnderscore{}＠\AgdaUnderscore{}}}\AgdaSymbol{)}\AgdaSpace{}%
\AgdaSymbol{(}\AgdaBound{A}\AgdaSpace{}%
\AgdaOperator{\AgdaBound{＠}}\AgdaSpace{}%
\AgdaBound{r}\AgdaSymbol{)}\<%
\\
\>[0]\AgdaBound{s}\AgdaSpace{}%
\AgdaOperator{\AgdaFunction{⇒}}\AgdaSpace{}%
\AgdaBound{r}\AgdaSpace{}%
\AgdaOperator{\AgdaFunction{◇}}\AgdaSpace{}%
\AgdaBound{t}\AgdaSpace{}%
\AgdaSymbol{=}\AgdaSpace{}%
\AgdaFunction{perform}\AgdaSpace{}%
\AgdaSymbol{(}\AgdaInductiveConstructor{`comm}\AgdaSpace{}%
\AgdaBound{s}\AgdaSpace{}%
\AgdaBound{r}\AgdaSpace{}%
\AgdaBound{t}\AgdaSymbol{)}\<%
\end{code}\end{center}
\end{minipage}

\begin{code}[hide]%
\>[0]\AgdaFunction{id-monad}\AgdaSpace{}%
\AgdaSymbol{:}\AgdaSpace{}%
\AgdaSymbol{∀}\AgdaSpace{}%
\AgdaSymbol{\{}\AgdaBound{ℓ}\AgdaSymbol{\}}\AgdaSpace{}%
\AgdaSymbol{→}\AgdaSpace{}%
\AgdaRecord{RawMonad}\AgdaSpace{}%
\AgdaSymbol{\{}\AgdaBound{ℓ}\AgdaSymbol{\}}\AgdaSpace{}%
\AgdaSymbol{(λ}\AgdaSpace{}%
\AgdaBound{A}\AgdaSpace{}%
\AgdaSymbol{→}\AgdaSpace{}%
\AgdaBound{A}\AgdaSymbol{)}\<%
\\
\>[0]\AgdaFunction{id-monad}\AgdaSpace{}%
\AgdaSymbol{=}\AgdaSpace{}%
\AgdaFunction{mkRawMonad}\AgdaSpace{}%
\AgdaSymbol{\AgdaUnderscore{}}\AgdaSpace{}%
\AgdaSymbol{(λ}\AgdaSpace{}%
\AgdaBound{x}\AgdaSpace{}%
\AgdaSymbol{→}\AgdaSpace{}%
\AgdaBound{x}\AgdaSymbol{)}\AgdaSpace{}%
\AgdaSymbol{(λ}\AgdaSpace{}%
\AgdaBound{x}\AgdaSpace{}%
\AgdaBound{f}\AgdaSpace{}%
\AgdaSymbol{→}\AgdaSpace{}%
\AgdaBound{f}\AgdaSpace{}%
\AgdaBound{x}\AgdaSymbol{)}\<%
\\
\\[\AgdaEmptyExtraSkip]%
\>[0]\AgdaFunction{top-monad}\AgdaSpace{}%
\AgdaSymbol{:}\AgdaSpace{}%
\AgdaSymbol{∀}\AgdaSpace{}%
\AgdaSymbol{\{}\AgdaBound{ℓ}\AgdaSpace{}%
\AgdaBound{ℓ′}\AgdaSymbol{\}}\AgdaSpace{}%
\AgdaSymbol{→}\AgdaSpace{}%
\AgdaRecord{RawMonad}\AgdaSpace{}%
\AgdaSymbol{\{}\AgdaBound{ℓ}\AgdaSymbol{\}}\AgdaSpace{}%
\AgdaSymbol{\{}\AgdaBound{ℓ′}\AgdaSymbol{\}}\AgdaSpace{}%
\AgdaSymbol{(λ}\AgdaSpace{}%
\AgdaBound{A}\AgdaSpace{}%
\AgdaSymbol{→}\AgdaSpace{}%
\AgdaRecord{⊤}\AgdaSymbol{)}\<%
\\
\>[0]\AgdaFunction{top-monad}\AgdaSpace{}%
\AgdaSymbol{=}\AgdaSpace{}%
\AgdaFunction{mkRawMonad}\AgdaSpace{}%
\AgdaSymbol{\AgdaUnderscore{}}\AgdaSpace{}%
\AgdaSymbol{(λ}\AgdaSpace{}%
\AgdaBound{\AgdaUnderscore{}}\AgdaSpace{}%
\AgdaSymbol{→}\AgdaSpace{}%
\AgdaInductiveConstructor{tt}\AgdaSymbol{)}\AgdaSpace{}%
\AgdaSymbol{(λ}\AgdaSpace{}%
\AgdaBound{\AgdaUnderscore{}}\AgdaSpace{}%
\AgdaBound{\AgdaUnderscore{}}\AgdaSpace{}%
\AgdaSymbol{→}\AgdaSpace{}%
\AgdaInductiveConstructor{tt}\AgdaSymbol{)}\<%
\\
\\[\AgdaEmptyExtraSkip]%
\>[0]\AgdaKeyword{instance}\<%
\\
\>[0][@{}l@{\AgdaIndent{0}}]%
\>[2]\AgdaFunction{focus-monad}\AgdaSpace{}%
\AgdaSymbol{:}\AgdaSpace{}%
\AgdaSymbol{∀}\AgdaSpace{}%
\AgdaSymbol{\{}\AgdaBound{ℓ}\AgdaSymbol{\}}\AgdaSpace{}%
\AgdaSymbol{\{}\AgdaBound{l}\AgdaSpace{}%
\AgdaBound{s}\AgdaSymbol{\}}\AgdaSpace{}%
\AgdaSymbol{→}\AgdaSpace{}%
\AgdaRecord{RawMonad}\AgdaSpace{}%
\AgdaSymbol{\{}\AgdaBound{ℓ}\AgdaSymbol{\}}\AgdaSpace{}%
\AgdaSymbol{(λ}\AgdaSpace{}%
\AgdaBound{A}\AgdaSpace{}%
\AgdaSymbol{→}\AgdaSpace{}%
\AgdaFunction{focus}\AgdaSpace{}%
\AgdaBound{l}\AgdaSpace{}%
\AgdaBound{A}\AgdaSpace{}%
\AgdaBound{s}\AgdaSymbol{)}\<%
\\
\>[2]\AgdaFunction{focus-monad}\AgdaSpace{}%
\AgdaSymbol{\{}\AgdaArgument{l}\AgdaSpace{}%
\AgdaSymbol{=}\AgdaSpace{}%
\AgdaBound{l}\AgdaSymbol{\}}\AgdaSpace{}%
\AgdaSymbol{\{}\AgdaArgument{s}\AgdaSpace{}%
\AgdaSymbol{=}\AgdaSpace{}%
\AgdaBound{s}\AgdaSymbol{\}}\AgdaSpace{}%
\AgdaKeyword{with}\AgdaSpace{}%
\AgdaBound{l}\AgdaSpace{}%
\AgdaOperator{\AgdaFunction{≟}}\AgdaSpace{}%
\AgdaBound{s}\<%
\\
\>[2]\AgdaSymbol{...}\AgdaSpace{}%
\AgdaSymbol{|}\AgdaSpace{}%
\AgdaInductiveConstructor{yes}\AgdaSpace{}%
\AgdaSymbol{\AgdaUnderscore{}}\AgdaSpace{}%
\AgdaSymbol{=}\AgdaSpace{}%
\AgdaFunction{id-monad}\<%
\\
\>[2]\AgdaSymbol{...}\AgdaSpace{}%
\AgdaSymbol{|}\AgdaSpace{}%
\AgdaInductiveConstructor{no}%
\>[12]\AgdaSymbol{\AgdaUnderscore{}}\AgdaSpace{}%
\AgdaSymbol{=}\AgdaSpace{}%
\AgdaFunction{top-monad}\<%
\end{code}

\caption{Choreographies as Algebraic Effects}
\label{fig:choreo}
\end{figure}

\Cref{fig:choreo} presents choreographies as algebraic effects.
One issue that every library-level CP language needs to deal with is how to represent located values.
Located values are variables in a choreography that denote values at different locations.
We give them types $A \ @ \ l$, which intuitively means a value of type $A$ at location $l$.

Existing library-level CP languages such as HasChor~\citep{shen-2023} define located values as a union of a plain value and a unit value --- an option type --- and have the unspoken invariant that when projecting to location $l$, values at $l$ are a plain value, and otherwise, a unit value.
Internally, HasChor uses an unwrap function to extract the plain value from the union.
The unwrap function is non-total because the union could be a unit value, but HasChor meticulously use it only in situations where the union is a plain value (implicitly use the invariant), so the non-totality never shows up.
This approach does not work in Agda, as it demands that every function be total.
For this reason, we take an alternative approach to located values, in which they are kept abstract and erased before projection in a way that respects the invariant by construction.
We first define \AgdaFunction{At}, a type-level function that captures the interface of located values.
Then, we define \AgdaFunction{focus}, a particular \AgdaFunction{At} that we will use in endpoint projection (we will show another \AgdaFunction{At} in the next section).
Intuitively, \AgdaFunction{focus} $l$ erases a located value of type $A \ @ \ s$ to $A$ if $l$ is equal to $s$; otherwise, to a unit value.

The signature ℂ specifies the two main operations of choreographies using one overloaded constructor:
\begin{itemize}
\item
  \AgdaInductiveConstructor{`comm} $s$ $s$ $t$ denotes a local computation $t$ at location $s$.
\item
  \AgdaInductiveConstructor{`comm} $s$ $r$ $t$ denotes location $s$ sends the result of a computation $t$ to location $r$.
\end{itemize}

We also define \AgdaFunction{Choreo} as a shorthand for terms using operations from ℂ abstracted over a particular \AgdaFunction{At}.
We also require \AgdaFunction{At} to be an instance of monads for any location $l$, which allows us to chain together located values.
Our focus is a monad because the identity functor and units are both monads.
We also define two helpful functions, \AgdaFunction{\_▷\_} and \AgdaFunction{\_⇒\_◇\_}, for writing choreographies.

\subsection{Endpoint Projection} \label{sec:epp}

\begin{code}[hide]%
\>[0]\AgdaKeyword{open}\AgdaSpace{}%
\AgdaModule{Process}\<%
\end{code}

\begin{figure}[ht]
\begin{code}%
\>[0]\AgdaFunction{epp}\AgdaSpace{}%
\AgdaSymbol{:}\AgdaSpace{}%
\AgdaSymbol{∀}\AgdaSpace{}%
\AgdaSymbol{\{}\AgdaBound{F}\AgdaSymbol{\}}\AgdaSpace{}%
\AgdaSymbol{→}\AgdaSpace{}%
\AgdaFunction{Choreo}\AgdaSpace{}%
\AgdaBound{F}\AgdaSpace{}%
\AgdaSymbol{→}\AgdaSpace{}%
\AgdaSymbol{(}\AgdaBound{l}\AgdaSpace{}%
\AgdaSymbol{:}\AgdaSpace{}%
\AgdaFunction{Loc}\AgdaSymbol{)}\AgdaSpace{}%
\AgdaSymbol{→}\AgdaSpace{}%
\AgdaFunction{Process}\AgdaSpace{}%
\AgdaSymbol{(}\AgdaBound{F}\AgdaSpace{}%
\AgdaSymbol{(}\AgdaFunction{focus}\AgdaSpace{}%
\AgdaBound{l}\AgdaSymbol{))}\<%
\\
\>[0]\AgdaFunction{epp}\AgdaSpace{}%
\AgdaBound{c}\AgdaSpace{}%
\AgdaBound{l}\AgdaSpace{}%
\AgdaSymbol{=}\AgdaSpace{}%
\AgdaFunction{interp}\AgdaSpace{}%
\AgdaFunction{alg}\AgdaSpace{}%
\AgdaFunction{return}\AgdaSpace{}%
\AgdaBound{c}\<%
\\
\>[0][@{}l@{\AgdaIndent{0}}]%
\>[2]\AgdaKeyword{where}\<%
\\
\>[2][@{}l@{\AgdaIndent{0}}]%
\>[4]\AgdaFunction{alg}\AgdaSpace{}%
\AgdaSymbol{:}\AgdaSpace{}%
\AgdaSymbol{∀}\AgdaSpace{}%
\AgdaSymbol{\{}\AgdaBound{A}\AgdaSymbol{\}}\AgdaSpace{}%
\AgdaSymbol{→}\AgdaSpace{}%
\AgdaFunction{ℂ}\AgdaSpace{}%
\AgdaSymbol{(}\AgdaFunction{focus}\AgdaSpace{}%
\AgdaBound{l}\AgdaSymbol{)}\AgdaSpace{}%
\AgdaOperator{\AgdaFunction{-Alg[}}\AgdaSpace{}%
\AgdaFunction{Process}\AgdaSpace{}%
\AgdaBound{A}\AgdaSpace{}%
\AgdaOperator{\AgdaFunction{]}}\<%
\\
\>[4]\AgdaFunction{alg}\AgdaSpace{}%
\AgdaSymbol{(}\AgdaInductiveConstructor{`comm}\AgdaSpace{}%
\AgdaBound{s}\AgdaSpace{}%
\AgdaBound{r}\AgdaSpace{}%
\AgdaBound{a}\AgdaSpace{}%
\AgdaOperator{\AgdaInductiveConstructor{,}}\AgdaSpace{}%
\AgdaBound{k}\AgdaSymbol{)}\AgdaSpace{}%
\AgdaKeyword{with}\AgdaSpace{}%
\AgdaBound{l}\AgdaSpace{}%
\AgdaOperator{\AgdaFunction{≟}}\AgdaSpace{}%
\AgdaBound{s}\AgdaSpace{}%
\AgdaSymbol{|}\AgdaSpace{}%
\AgdaBound{l}\AgdaSpace{}%
\AgdaOperator{\AgdaFunction{≟}}\AgdaSpace{}%
\AgdaBound{r}\<%
\\
\>[4]\AgdaSymbol{...}\AgdaSpace{}%
\AgdaSymbol{|}\AgdaSpace{}%
\AgdaInductiveConstructor{yes}\AgdaSpace{}%
\AgdaSymbol{\AgdaUnderscore{}}%
\>[17]\AgdaSymbol{|}\AgdaSpace{}%
\AgdaInductiveConstructor{yes}\AgdaSpace{}%
\AgdaSymbol{\AgdaUnderscore{}}%
\>[26]\AgdaSymbol{=}\AgdaSpace{}%
\AgdaFunction{locally}\AgdaSpace{}%
\AgdaBound{a}\AgdaSpace{}%
\AgdaOperator{\AgdaField{>>=}}\AgdaSpace{}%
\AgdaBound{k}\<%
\\
\>[4]\AgdaSymbol{...}\AgdaSpace{}%
\AgdaSymbol{|}\AgdaSpace{}%
\AgdaInductiveConstructor{yes}\AgdaSpace{}%
\AgdaSymbol{\AgdaUnderscore{}}%
\>[17]\AgdaSymbol{|}\AgdaSpace{}%
\AgdaInductiveConstructor{no}%
\>[23]\AgdaSymbol{\AgdaUnderscore{}}%
\>[26]\AgdaSymbol{=}\AgdaSpace{}%
\AgdaFunction{locally}\AgdaSpace{}%
\AgdaBound{a}\AgdaSpace{}%
\AgdaOperator{\AgdaField{>>=}}\AgdaSpace{}%
\AgdaSymbol{(λ}\AgdaSpace{}%
\AgdaBound{x}\AgdaSpace{}%
\AgdaSymbol{→}\AgdaSpace{}%
\AgdaFunction{send}\AgdaSpace{}%
\AgdaBound{r}\AgdaSpace{}%
\AgdaBound{x}\AgdaSymbol{)}\AgdaSpace{}%
\AgdaOperator{\AgdaFunction{>>}}\AgdaSpace{}%
\AgdaBound{k}\AgdaSpace{}%
\AgdaInductiveConstructor{tt}\<%
\\
\>[4]\AgdaSymbol{...}\AgdaSpace{}%
\AgdaSymbol{|}\AgdaSpace{}%
\AgdaInductiveConstructor{no}%
\>[14]\AgdaSymbol{\AgdaUnderscore{}}%
\>[17]\AgdaSymbol{|}\AgdaSpace{}%
\AgdaInductiveConstructor{yes}\AgdaSpace{}%
\AgdaSymbol{\AgdaUnderscore{}}%
\>[26]\AgdaSymbol{=}\AgdaSpace{}%
\AgdaFunction{recv}\AgdaSpace{}%
\AgdaBound{s}\AgdaSpace{}%
\AgdaOperator{\AgdaField{>>=}}\AgdaSpace{}%
\AgdaBound{k}\<%
\\
\>[4]\AgdaSymbol{...}\AgdaSpace{}%
\AgdaSymbol{|}\AgdaSpace{}%
\AgdaInductiveConstructor{no}%
\>[14]\AgdaSymbol{\AgdaUnderscore{}}%
\>[17]\AgdaSymbol{|}\AgdaSpace{}%
\AgdaInductiveConstructor{no}%
\>[23]\AgdaSymbol{\AgdaUnderscore{}}%
\>[26]\AgdaSymbol{=}\AgdaSpace{}%
\AgdaBound{k}\AgdaSpace{}%
\AgdaInductiveConstructor{tt}\<%
\end{code}
\caption{Endpoint Projection}
\label{fig:epp}
\end{figure}

We can now define endpoint projection, the process of turning a choreography into a process for a target location.
\Cref{fig:epp} presents our implementation of EPP.
The function \AgdaFunction{epp} takes a choreography $c$ and a target location $l$, and uses the effect handler \AgdaFunction{interp} to interpret operations in $c$.
For variables, we return them in the generated process.
For operations, we construct a \AgdaFunction{ℂ}-Algebra (with all located values erased from $l$'s perspective) \AgdaFunction{alg} on processes, which does one step of interpretation.
The only operation we need to interpret is \AgdaInductiveConstructor{`comm}, depending on whether $l$ is equal to $s$ and $r$:
\begin{itemize}
\item
  If $l$ equals $s$ and $r$, meaning $s$ and $r$ are the same, we interpret this operation as a local computation followed by the continuation.
\item
  If $l$ equals $s$ but not $r$, meaning the target location is the sender, we interpret the operation as a local computation followed by a send and the continuation.
\item
  If $l$ equals $r$ but not $s$, meaning the target location is the receiver, we interpret the operation as a receive followed by the continuation.
\item
  If $l$ equals neither $s$ nor $r$, meaning the target location is not involved, we just return the continuation.
\end{itemize}

\section{Next Steps}
\label{sec:next}

\begin{code}[hide]%
\>[0]\AgdaKeyword{open}\AgdaSpace{}%
\AgdaKeyword{import}\AgdaSpace{}%
\AgdaModule{02-algeff}\<%
\\
\\[\AgdaEmptyExtraSkip]%
\>[0]\AgdaKeyword{module}\AgdaSpace{}%
\AgdaModule{05-next}\AgdaSpace{}%
\AgdaSymbol{(}\AgdaBound{𝕃}\AgdaSpace{}%
\AgdaSymbol{:}\AgdaSpace{}%
\AgdaRecord{Sig}\AgdaSymbol{)}\AgdaSpace{}%
\AgdaKeyword{where}\<%
\\
\\[\AgdaEmptyExtraSkip]%
\>[0]\AgdaKeyword{import}\AgdaSpace{}%
\AgdaModule{03-choreo}\<%
\\
\>[0]\AgdaKeyword{import}\AgdaSpace{}%
\AgdaModule{04-proofs}\<%
\\
\\[\AgdaEmptyExtraSkip]%
\>[0]\AgdaKeyword{open}\AgdaSpace{}%
\AgdaKeyword{module}\AgdaSpace{}%
\AgdaModule{03-choreo′}\AgdaSpace{}%
\AgdaSymbol{=}\AgdaSpace{}%
\AgdaModule{03-choreo}\AgdaSpace{}%
\AgdaBound{𝕃}\<%
\\
\>[0]\AgdaKeyword{open}\AgdaSpace{}%
\AgdaKeyword{module}\AgdaSpace{}%
\AgdaModule{04-proofs′}\AgdaSpace{}%
\AgdaSymbol{=}\AgdaSpace{}%
\AgdaModule{04-proofs}\AgdaSpace{}%
\AgdaBound{𝕃}\<%
\\
\>[0]\<%
\end{code}

Our next step will be to leverage the algebraic-effects-based formulation of CP presented in the last two sections to prove the correctness of endpoint projection, as well as follow-on properties such as deadlock freedom.

At a high level, given a choreographic semantics \AgdaDatatype{\_⇒ᶜ\_} and a network semantics \AgdaDatatype{\_⇒ⁿ\_}, \emph{soundness} of endpoint projection would say that the projected network \emph{preserves} the semantics of the original choreography.
That is, for choreographies $c$ and $c'$,
\begin{code}[hide]%
\>[0]\AgdaFunction{\AgdaUnderscore{}}%
\>[19I]\AgdaSymbol{=}\AgdaSpace{}%
\AgdaSymbol{∀}\AgdaSpace{}%
\AgdaSymbol{\{}\AgdaBound{F}\AgdaSymbol{\}}\AgdaSpace{}%
\AgdaSymbol{(}\AgdaBound{c}\AgdaSpace{}%
\AgdaBound{c′}\AgdaSpace{}%
\AgdaSymbol{:}\AgdaSpace{}%
\AgdaFunction{Choreo}\AgdaSpace{}%
\AgdaBound{F}\AgdaSymbol{)}\AgdaSpace{}%
\AgdaSymbol{→}\<%
\end{code}
\begin{code}[inline]%
\>[.][@{}l@{}]\<[19I]%
\>[2]\AgdaFunction{epp}\AgdaSpace{}%
\AgdaBound{c}\AgdaSpace{}%
\AgdaOperator{\AgdaDatatype{⇒ⁿ}}\AgdaSpace{}%
\AgdaFunction{epp}\AgdaSpace{}%
\AgdaBound{c′}\AgdaSpace{}%
\AgdaSymbol{→}\AgdaSpace{}%
\AgdaBound{c}\AgdaSpace{}%
\AgdaOperator{\AgdaDatatype{⇒ᶜ}}\AgdaSpace{}%
\AgdaBound{c′}\<%
\end{code}.
\emph{Completeness} of EPP, on the other hand, would say that the network \emph{reflects} the semantics of the original choreography, that is, given choreographies $c$ and $c'$,
\begin{code}[hide]%
\>[0]\AgdaFunction{\AgdaUnderscore{}}%
\>[36I]\AgdaSymbol{=}\AgdaSpace{}%
\AgdaSymbol{∀}\AgdaSpace{}%
\AgdaSymbol{\{}\AgdaBound{F}\AgdaSymbol{\}}\AgdaSpace{}%
\AgdaSymbol{(}\AgdaBound{c}\AgdaSpace{}%
\AgdaBound{c′}\AgdaSpace{}%
\AgdaSymbol{:}\AgdaSpace{}%
\AgdaFunction{Choreo}\AgdaSpace{}%
\AgdaBound{F}\AgdaSymbol{)}\AgdaSpace{}%
\AgdaSymbol{→}\<%
\end{code}
\begin{code}[inline]%
\>[.][@{}l@{}]\<[36I]%
\>[2]\AgdaBound{c}\AgdaSpace{}%
\AgdaOperator{\AgdaDatatype{⇒ᶜ}}\AgdaSpace{}%
\AgdaBound{c′}\AgdaSpace{}%
\AgdaSymbol{→}\AgdaSpace{}%
\AgdaFunction{epp}\AgdaSpace{}%
\AgdaBound{c}\AgdaSpace{}%
\AgdaOperator{\AgdaDatatype{⇒ⁿ}}\AgdaSpace{}%
\AgdaFunction{epp}\AgdaSpace{}%
\AgdaBound{c′}\<%
\end{code}.
If these correctness conditions hold, the transition systems \AgdaDatatype{\_⇒ᶜ\_} and \AgdaDatatype{\_⇒ⁿ\_} are in bisimulation,
and if choreographies enjoy a progress property, networks also have it, implying that they are deadlock-free.
However, the above definitions of soundness and completeness may be too strong --- for instance, they prohibit EPP from doing rewritings that change the behaviors of the network but compute the same result.
Thus, we are working on more relaxed correctness conditions that permit more interesting behaviors in networks while still maintaining deadlock freedom.

In the longer term, we want to bring our algebraic-effects-based formulation of CP to languages with efficient native support for algebraic effects, such as OCaml and Koka.

\bibliographystyle{ACM-Reference-Format}
\bibliography{10-references}

\end{document}